\documentclass[journal=jacsat,manuscript=article]{achemso}

\usepackage{chemformula} 
\usepackage[T1]{fontenc} 
\usepackage{graphicx}
\usepackage{dcolumn}
\usepackage{bm}
\usepackage{mathtools}
\usepackage{mathrsfs, amsmath, wasysym}
\usepackage[mathscr]{eucal}
\usepackage{xfrac}
\usepackage{graphicx}
\usepackage{dcolumn}
\usepackage{bm}
\usepackage{color}
\usepackage{tabularx}
\usepackage{hyperref}
\hypersetup{colorlinks,linkcolor=blue,citecolor=blue,urlcolor=blue}
\usepackage[normalem]{ulem}
\usepackage[all]{hypcap}

\author{Zhenyao Fang}
\affiliation[Northeastern University]{Department of Physics, Northeastern University, Boston, MA 02115, USA}
\email{z.fang@northeastern.edu}
\author{Qimin Yan}
\affiliation[Northeastern University]{Department of Physics, Northeastern University, Boston, MA 02115, USA}
\email{q.yan@northeastern.edu}

\title{Leveraging Persistent Homology Features for Accurate Defect Formation Energy Predictions via Graph Neural Networks}
\abbreviations{persistent homology features, defect properties, graph neural networks}

\begin{document}

\begin{abstract}
In machine-learning-assisted high-throughput defect studies, a defect-aware latent representation of the supercell structure is crucial to the accurate prediction of defect properties. The performance of current graph neural network (GNN) models is limited due to the fact that defect properties depend strongly on the local atomic configurations near the defect sites and due to the over-smoothing problem of GNN. Herein, we demonstrate that persistent homology features, which encode the topological information of local chemical environment around each atomic site, can characterize the structural information of defects. Using the dataset containing a wide spectrum of \ch{O}-based perovskites with all available vacancies as an example, we show that incorporating the persistent homology features, along with proper choices of graph pooling operations, significantly increases the prediction accuracy, with the MAE reduced by 55\%. Those features can be easily integrated into the state-of-the-art GNN models, including the graph Transformer network and the equivariant neural network, and universally improve their performance. Besides, our model also overcomes the convergence issue with respect to the supercell size that was present in previous GNN models. Furthermore, using the datasets of defective \ch{BaTiO3} with multiple substitutions and multiple vacancies as examples, our GNN model can also predict the defect-defect interactions accurately. These results suggest that persistent homology features can effectively improve the performance of machine learning models and assist the accelerated discovery of functional defects for technological applications.
\end{abstract}

\section{Introduction}
Understanding the effect of point defects in materials has been a central area of study in materials science, since defects can significantly influence the physical and chemical properties of materials (such as reducing the bulk modulus and inducing structural phase transitions~\cite{Li22mechanical_review, Bostrom21mechanical_review, Dastider23mechanical_review}), and introducing in-gap defect states which can further affect their performance in optoelectronic devices and catalysts~\cite{Pan13electronic_review, Yin15electronic_review, Leem24electronic_review}. One of the key quantities in defect chemistry is the defect formation energy~\cite{VdW04defect_review, Freysoldt14defect_review}, which can affect various defect properties, such as the favorable type of defects and the location of defects in a given host material, the defect concentration at certain temperature, and defect migration pathways near surfaces or grain boundaries \cite{Oba11defect_migration, Pochet12defect_migration}. Therefore, fast and accurate predictions of defect formation energies are crucial to understanding the defect-related phenomenon in materials, enabling high-throughput screening of host material with desired defect properties.

To evaluate the defect formation energies, first-principles calculations based on the density functional theory (DFT) are often employed~\cite{Freysoldt14defect_review, Lany08defect_DFT, Freysoldt18defect_DFT}. However, this method often involves large supercell calculations that are computationally expensive. Therefore, various machine learning methods, along with physics-inspired descriptors (such as the formation enthalpy, the energy above hull, the band gap, and the crystal reduction potential), were proposed to accelerate the predictions of defect formation energies~\cite{Frey20ML, Deml15ML, Wexler21ML, Park21ML, Yan24ML}. Although these machine learning models have achieved great accuracy in predicting defect formation energies, they lack the generalizability to accommodate a wide spectrum of materials because they rely on human-selected features, which are designed specifically for certain types of materials. Besides, due to the model architecture, most reported (shallow) machine learning models cannot incorporate the local structural information and thus cannot evaluate the defect formation energies for different symmetry-inequivalent atomic sites with structural relaxations, restricting their applications in predicting defect formation energies in complex materials that can potentially host multiple defects.

Recent years, graph neural networks (GNNs) have shown remarkable success in predicting various material properties, including the formation enthalpy, band gap, and bulk modulus~\cite{Chen19GNN, Reiser22GNN, Fung21GNN}. In a GNN model, the input crystal structure is first converted into a graph, where the nodes and edges represent the atoms and bonds in the unit cell~\cite{Xie18CGCNN}. The graph then passes through several convolution layers, where the node information exchanges between adjacent nodes and aggregate together, allowing the GNN model to learn the local chemical environment of each atom and to extract latent representations of the whole crystal structure, which will be used to make predictions on material properties. Until recently, few GNN models have been applied to predict defect formation energies~\cite{Kazeev23Defect_GNN, Xiang24Defect_GNN, Rahman24Defect_ALIGNN}. As first attempts to use GNNs to predict defect related properties, these models incorporated various levels of chemical information, including the bond angle which may help detect the disrupted chemical bonding network near the vacancies \cite{Rahman24Defect_ALIGNN}, but their prediction accuracy is currently limited for a diverse set of materials. The reasons for the poor performance of GNNs on predicting defect formation energies are two-fold. Firstly, defect formation energies highly depend on the atomic configurations near the defective sites and are independent of the atoms far from the defective sites, and are thus local properties of the crystal, contrary to the formation enthalpy or the band gap that are global quantities of the whole crystal. Secondly, GNNs suffer from the over-smoothing problem~\cite{Li18oversmoothing_GNN}. When passing through consecutive convolution layers, features between adjacent nodes will exchange and aggregate, and finally converge to constant values in a deep GNN model. This could lead to incorrect predictions especially when the prediction target only depends on the local atomic configuration. Therefore, in order for GNNs to accurately predict the defect formation energies, the structural information about the defects, such as the distance to the defect site or the number of nearby defects, must be encoded explicitly in the features of each node. This approach amplifies the differences in node features between the nodes near and far from the defect site, which can potentially reduce the over-smoothing effect of GNN and direct the focus of the neural network toward the nodes near the defects.

Based on these motivations, we propose to incorporate the persistent homology features into GNN models to predict the defect formation energy accurately. Persistent homology features are generated from the topological invariants of the homology group associated with the surrounding atomic configurations near each atom~\cite{Pun22persistent_homology, Wu18persistent_homology}. These features have already been applied in topological data analysis and in predicting (global) properties of molecules and materials, such as the formation enthalpy~\cite{Cang17persistent_homology, Jiang21persistent_homology}. In this work, we demonstrate that persistent homology features encode the local structural information related to defects and can significantly increase the prediction accuracy of defect formation energies in complex materials. We show from several toy models that persistent homology features contain such structural information as the type and the size of defects, the number of defects around each atom (within a cutoff radius), and the distance to the defects from each atom. To verify the efficacy of persistent homology features, we train various GNN models on the dataset containing around 7700 \ch{O}-based perovskite defective structures with mono-vacancies to predict their neutral defect formation energies. We find that introducing the persistent homology features significantly increases the performance of GNN models, with the mean absolute error (MAE) reduced by 55\%. Besides, our model with the persistent homology features and the global max pooling layers can overcome the convergence issue with respect to the supercell size reported previously~\cite{Rahman24Defect_ALIGNN}. Furthermore, we also apply the GNN model to datasets of a prototypical cubic perovskite \ch{BaTiO3}~\cite{Roedel09BaTiO3} with multiple substitutions and multiple vacancies. We show that the persistent homology features also help capturing the defect-defect interaction energies in those datasets. These results suggest the crucial role of persistent homology features in extracting local structural information near defects and in predicting physical and chemical properties related to defects in complex materials using machine learning or deep learning methods.

\section{Results}
\subsection{Persistent Homology Features and Defects} \label{sec:ASPH}
A simplicial complex $X$ is formed by gluing "standard" geometric objects (simplicies) of various dimensions (such as points, line segments, triangles, tetrahedrons) together. One of its homological properties is the $d$-dimensional Betti number $\beta_d (X)$, defined as the rank of the homology group $H_d(X)$. Any nontrivial element in the homology group $H_d(X)$ corresponds to some $d$-dimensional cycle (cyclic simplices) that is not simultaneously the boundary of any $(d+1)$-dimensional chain, namely a $d$-dimensional "hole"; for example, a circle is a one-dimensional cycle and encloses a hole, whereas the boundary of a disk, though also being a one-dimensional cycle, does not enclose a hole since it is the boundary of the two-dimensional disk. Therefore, the $d$-Betti numbers characterize the number of $d$-dimensional holes in the simplicial complex and reflect the connectivity of the simplicial complex (see Supplementary Materials I for definitions)~\cite{dey22persistent_homology}. In the context of topological data analysis, any set of data points can be converted into a simplicial complex by connecting the points within some cutoff radius together. The cutoff radius is allowed to vary (to an upper bound) so that the underlying pattern behind the given set of data points can be revealed through the birth, the death, and the persistence of Betti numbers (see Supplementary Materials I for more discussions)~\cite{Pun22persistent_homology, Aktas19persistent_homology, Zia24persistent_homology}. 

In chemical systems such as molecules or materials, the atomic configuration near each atom can be naturally regarded as a point cloud, and the chemical bonding network as a simplicial complex, allowing for the calculations of persistent homology features. Previous work introduced atom-specific persistent homology (ASPH) features as a topological representation of the crystal structure~\cite{Jiang21persistent_homology}. For each base atom and each chemical species in the unit cell, we generate a point cloud which centers at the base atom and includes all other atoms of the given chemical species within the given upper bound cutoff radius. The persistent homology features are then calculated for each point cloud; since crystalline materials are three-dimensional, only 0-, 1-, and 2-Betti numbers are relevant. The birth, death, and persistence of these Betti number features are further characterized by five statistical quantities, including the minimum, maximum, mean, standard deviation, and the (weighted) sum, resulting in 35 statistical representations of the persistent homology features for each atom in the unit cell (for 0-Betti numbers, only its death will be considered). An example is shown in Figure~1(A), where the crossed atom represents the vacancy in the two-dimensional lattice. The point cloud centered at the red atom includes the vacancy site, while that centered at the black atom excludes the vacancy site. The 0- and 1-Betti numbers are calculated for these two point clouds and their statistical quantities will be collected as the persistent homology features for these two atoms, respectively.

\begin{figure}[htb]
\centering
\includegraphics[width=\linewidth]{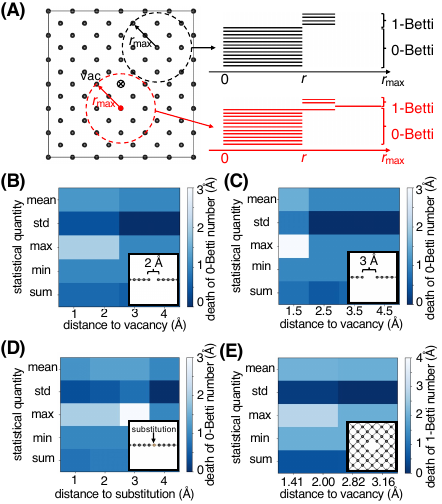}
\caption{(A) A schematic plot of the calculation procedure of atoms-specific persistent homology features. (B) The calculated statistical quantity of the death of 0-Betti number features for the one-dimensional toy model with one vacancy. (C, D) Same as (B), but for the toy model with two adjacent vacancies (C) and with a substitution (D) respectively. (E) The calculated statistical quantity of the death of the 1-Betti number features for the two-dimensional toy model with one vacancy.}
\label{fig:betti_numbers}
\end{figure}

Since persistent homology features contain information about holes within a point cloud, we conjecture that these features can unveil the structural changes induced by defects in materials. In this work we examine two of the most common defect types, namely vacancies and substitutions. Vacancies can be regarded as holes in a material and thus be detected by persistent homology features, while substitutions, as we will demonstrate below, are also encoded in these ASPH features because the point cloud is constructed for each chemical species around the base atom. 

To verify our hypothesis, we construct a toy model consisting of a one-dimensional atomic chain with equal interatomic spacing (1~\AA), and we create a vacancy in the chain, as shown in Figure~1(B). Since this is a one-dimensional chain, only the 0-Betti number features are relevant. Therefore, we calculate the 0-Betti number features for all atoms with the upper bound cutoff radius to be 3~\AA~and generate the statistical quantities of them. As discussed in Supplementary Materials I, 0-Betti numbers reflect the connected components in the simplicial complex. For atoms that are at least 3~\AA~far away from the vacancy, the point cloud around them does not contain the vacancy site. Thus, the maximum death of their 0-Betti number features is 1~\AA, above which all atoms in the point cloud are connected with each other and form one trivial connected component. On the other hand, for atoms less than 3~\AA~away from the vacancy site, the maximum death of their 0-Betti number features increases to 2~\AA~which is required for the two connected components to the left and right of the vacancy to connect with each other. Therefore, the 0-Betti number features can help distinguish the atoms close to or far from the defect site, as determined by the upper bound cutoff radius.

Furthermore, we consider the toy model with two adjacent vacancies, and the calculated features are shown in Figure~1(C). In this case, we define the position of the vacancy to be the center of the two missing atoms, so the distance of the nearest neighbor to the vacancy site is 1.5~\AA. The maximum death of the 0-Betti number features for the atoms nearest to the vacancy site now increases to 3~\AA, because the cutoff radius must be at least 3~\AA~in order to connect all atoms together into one connected component, suggesting that 0-Betti number features contain information about the size of vacancies. On the other hand, the maximum death of the next nearest neighbor decreases to 1~\AA, since its surrounding point cloud can no longer contain any atom in the other side of the vacancy.

Apart from vacancies, we also argue that ASPH features are sensitive to substitutions. By definition, for each atom in the unit cell, the point cloud which will be used to calculate persistent homology features is constructed for each chemical species. Therefore, when an atom is substituted by an exotic species, the point cloud for this species will contain only two points (the base atom and the substituted atom), and the critical distance at which these two points are connected is exactly the distance to the substitution site. This is confirmed by the above toy model with one substitution, as shown in Figure~1(D). The calculated features clearly reflect the distance to the substitution from each atom.

Finally, in the two-dimensional case, not only the 0-Betti numbers, but also the 1-Betti numbers are relevant. Here we calculate the 1-Betti number properties for a two-dimensional lattice with one vacancy, shown in Figure~1(E). In this case, the death of the 1-Betti number indicates the critical cutoff radius at which the hole formed by the vacancy is filled, with the critical cutoff radius to be the diagonal length of the square. More discussions on other toy models, including two distant vacancies and multiple element cases, can be found in Supplementary Materials I. We anticipate that in three-dimensional complex materials with multiple elements and varying bond lengths and bond angles, 0-, 1-, and 2-Betti numbers are all nontrivial, and their statistical quantities can jointly capture the structural information near the defects, including the number of defects near each atom, the distance to those defects, and the size of the vacancies.

\subsection{Vacancy Formation Energy in \ch{O}-Based Perovskites}
To numerically verify the capability of persistent homology features in capturing local structural information of defects and improving the prediction accuracy of defect-related properties in complex materials, we construct GNN models to predict the neutral defect formation energies for various datasets according to Equation~\ref{eqn:defect_formation_energy}. These models can be easily adapted to predicting other defect properties such as charge transition levels and charged defect formation energies.

Our GNN model features a typical model architecture for graph property prediction, as shown in Figure~2. First, the defective supercell structure is converted into a graph, where the nodes and the edges of the graph represent the atoms and the chemical bonds in the crystal. The node features carry the atomic information of each atom, including the elemental features and the ASPH features. In this work, we consider three types of elemental features: atomic number (only), elemental properties (including the atomic number, atomic mass, atomic radius, electron affinity, row and group numbers), and the one-hot encoding of the elemental type~\cite{Xie18CGCNN}. The complexity and representation capability progressively increase for these three different options; the atomic number feature has been used to predict configurational disorder related properties, which usually involve only few elements~\cite{Fang24configurational_GNN}, while the one-hot encoding feature was used to predict the formation enthalpies and the band gaps of a wide range of materials in Materials Project \cite{Jain13Materials_project}, thus used in our work (benchmark results on elemental features can be found in Supplementary Materials II). On the other hand, the edge features represent the bonding properties, which in our case are simply the bond length expanded in a Gaussian function basis~\cite{Xie18CGCNN}. Note that in other types of higher-order GNN models, more bonding information of the crystal, such as the bond direction and the bond angle as well as the multiplet interactions, can also be included in the graph~\cite{Morris18higher_order_GNN, Reiser22higher_order_GNN}. 

The generated graph then passes through several graph convolution layers, where each node receives information from adjacent nodes and aggregate them together. Depending on the methods to aggregate and update node features, various convolution layers were proposed previously. In this work, we consider two convolution layers with the attention mechanism (the graph attention neural network (GAT)~\cite{GAT1, GAT2} and the Transformer network~\cite{Shi21Transformer}) and one without the attention mechanism (the crystal graph neural network (CGCNN)~\cite{Xie18CGCNN}). In attention-based neural networks, the importance of an adjacent node is expressed as the attention coefficient which is calculated from both the node features of the two nodes and edge features and reflects the similarity between the two node features (elemental properties and the persistent homology features) in the latent space. The node features are then updated by a linear combination of all adjacent node features weighted by the attention coefficients. The attention mechanism allows the GNN model to capture the chemical and structural distinctions between adjacent atoms and in general leads to better performance on various tasks~\cite{Zhou2020GNN_review}. In addition to those networks, we also consider the equivariant neural network (E3NN)~\cite{Batzner22EquivariantGNN}; this network employs the E(3)-equivariant convolution operation which utilizes the spatial symmetries in crystals to make accurate predictions on material properties. In the following, the convolution layer type is considered a hyperparameter and optimized through Bayesian optimization (see Methods section), unless otherwise specified. Finally, after the graph convolution layers, all node features in the graph are gathered together through graph pooling layers to produce the latent representation of the whole crystal, which will be further passed to a multilayer perceptron network to obtain the defect formation energy. We will demonstrate below that the choices of node features, graph convolution layers, and pooling layers significantly affect the performance of GNN models; for the tasks related to predicting defect properties, the optimal choices of the model architecture are different from typical GNN models used to predict the formation enthalpies or band gaps of bulk materials.

\begin{figure}[htb]
\centering
\includegraphics[width=\linewidth]{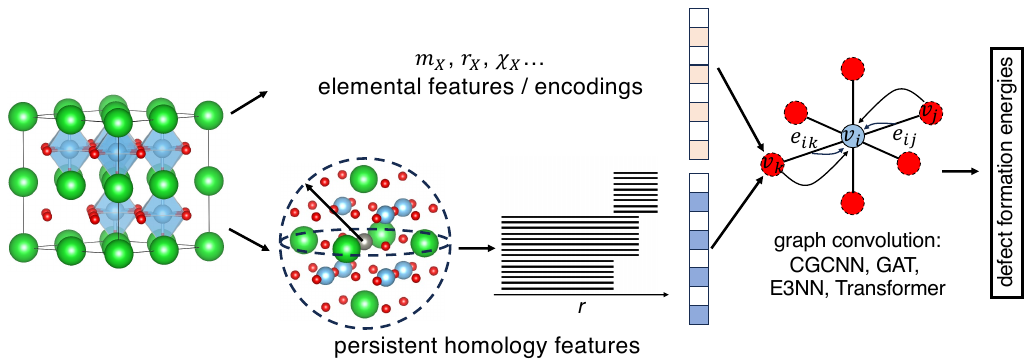}
\caption{A schematic plot of the GNN model based on the attention mechanism to predict the defect formation energies. The input defective structure will be converted into a crystal graph, where node features contain both the elemental features (such as the mass $m_X$, the radius $r_X$, and the electronegativity $\chi_X$ of element $X$) and the ASPH features, calculated by constructing the point cloud around each atom and tracking the birth/death/persistence of the homology features. The graph passes through several graph convolution layers, where features on adjacent nodes exchange and aggregate. The convoluted node features are gathered together into the global latent representation of the graph which will be used to predict the defect formation energy.}
\label{fig:GNN_architecture}
\end{figure}

Now we demonstrate the efficacy of persistent homology features using the dataset containing \ch{O}-based perovskite defective structures. Here we consider all available mono-vacancies in these structures. To construct our dataset, we choose all perovskite structures with the chemical formula \ch{ABO3} from Materials Project~\cite{Jain13Materials_project}, excluding those with cations after \ch{La} in the periodic table. Most elements after \ch{La} (such as the $f$-block elements, $6p$ and $6d$ elements) require further detailed investigations on the optimal $U(J)$ values that highly depend on individual perovskite structures within the DFT+U framework. Besides, excluding structures with those elements in our dataset does not affect the generalizability of our GNN models and the overall conclusion of our work. Next we use the pymatgen-analysis-defects package~\cite{Alkauskas14pymatgen_defect_analysis, Freysoldt09pymatgen_defect_analysis} to generate the supercells to calculate the defect formation energies for each perovskite structure. The supercells are constructed as nearly cubic cells with the minimum (maximum) lattice constants along each direction to be 10~\AA~(25~\AA), so that the interactions between defects and their periodic images along each Cartesian direction are minimal. Structures whose supercells cannot be constructed into the nearly-cubic cell within the size range are discarded. Using the generated supercells, we enumerate all available symmetry-inequivalent atomic sites and generate the corresponding vacancy structures (including both cation and anion vacancies). 

Using the above procedure, our dataset contains around 7700 defective structures derived from around 1100 host (pristine) structures. As shown in Figure 3(A), this dataset contains host structures from all crystal families, thus representing a comprehensive array of crystal structures. Besides, our dataset contains vacancies of all elemental types (before La), and the numbers of each type of vacancy are shown in Figure 3(B); some of the most common elemental types are also summarized in Table~1. The various atomic and chemical bonding configurations make it challenging for GNN models to learn the structural information and predict the defect formation energies.

\begin{figure}[htb]
\centering
\includegraphics[width=\linewidth]{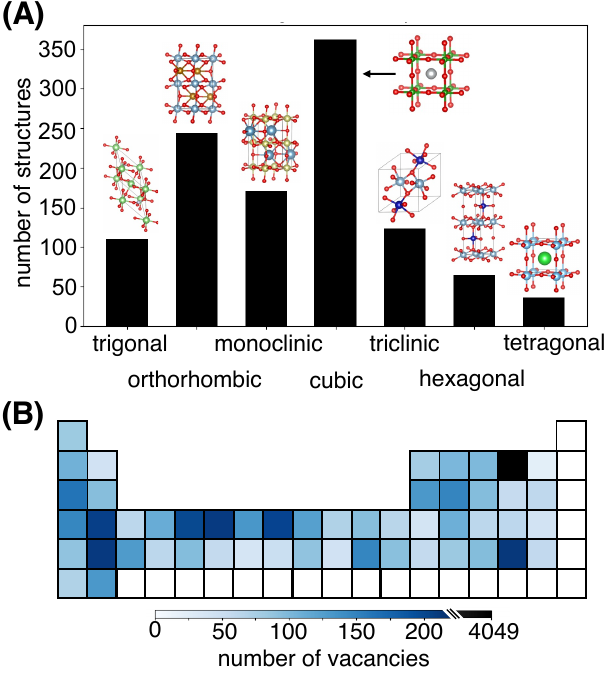}
\caption{(A) The number of host structures of each crystal family in our constructed \ch{O}-based perovskite dataset. The insets are example materials from each crystal family. (B) The number of vacancies of each elemental type in our dataset. Vacancies of all elemental types before \ch{La}, except for noble gas elements, are present in our dataset.}
\label{fig:dataset}
\end{figure}

\begin{table*}[htb]
    \centering
    \begin{tabular}{|c|c|c|c|c|c|c|c|}
    \hline
    vacancy type & \ch{O} & \ch{Sr} & \ch{Te} & \ch{Cr} & \ch{Ca} & \ch{Fe} & \ch{V} \\ \hline
    number & 4049 & 216 & 203 & 178 & 173 & 171 & 161 \\ \hline
    \end{tabular}
    \caption{The number of vacancies of the most common elemental types in the constructed \ch{O}-based perovskite dataset.}
    \label{tab:vacancy_type}
\end{table*}

First we present benchmark calculations on our constructed dataset using two state-of-the-art GNN models, namely the CGCNN model~\cite{Xie18CGCNN} and the higher-order model atomistic line GNN~\cite{Choudhary2021ALIGNN}. The latter model also incorporates the bonding angle using the line graph methods, allowing the model to capture the disrupted chemical bonding network near the defect sites and to make more accurate predictions about defect-related properties~\cite{Rahman24Defect_ALIGNN}. Both models used only the elemental features as the node features and global mean pooling layers. After training and testing on our dataset, the MAEs of these two GNN models are 1.86 and 1.83~eV, respectively. 

The poor performance of these models demonstrates the challenge of our constructed dataset, but more importantly, the uniqueness of defect property prediction tasks. Firstly, defect property calculations often require large supercell structures. Since GNN models involve graph pooling operations on all nodes and gather all node features as the global representation of the graph, the distinct node features of atoms near the defect site, generated by a disrupted chemical bonding network (as is the case in atomistic line GNN), are averaged out by the node features of the numerous atoms located far from the defects in the supercell. Therefore, the performance of GNN models generally worsens using larger supercells for defect calculations, and this trend was reported previously~\cite{Rahman24Defect_ALIGNN}. Secondly, GNN models suffer from the over-smoothing problems~\cite{Li18oversmoothing_GNN}. Since the node features are exchanged between adjacent nodes for each convolution layer, the node features will exponentially converge to constant values in a deep GNN model. In our scenario, this problem will also average out the node features and make the distinct features carried by atoms near the defect site invisible to the model. To overcome these problems, we choose to explicitly include the local chemical environment information associated with the defect site, represented as the ASPH features, into each node feature of the GNN model. These additional features can amplify the differences in node features between atoms located close and far from the defect sites. Besides, along with proper choices of graph pooling layers, these features are not averaged out by other atoms that do not contribute to the defect properties or by the over-smoothing effect, therefore enhancing the prediction accuracy of the GNN model.

For each defective structure, we generate the ASPH features (35 statistical quantities of the birth/death/persistence of Betti numbers) for each atom in the supercell~\cite{Jiang21persistent_homology}. When generating the features, the upper bound cutoff radius can affect the final performance of the GNN model. With larger upper bound cutoff radius, structural information from farther defects can be included, but computations will also become more expensive. In Supplementary Materials II, we present convergence test results on the upper bound cutoff radius, and decide to use 10~\AA~as the upper bound cutoff radius to generate the persistent homology features for the whole dataset. Besides, the generated homology features could be correlated with each other. To extract a concise feature vector and remove the redundant features, we pre-process the calculated ASPH features by the principal component analysis method, which reduces the dimensionality of features through linear transformations and concentrates the information in features into the principal components~\cite{Greenacre2023PCA}. As shown in Supplementary Materials II, the MAE reduces as more principal components are included into the node features and reaches convergence at around 6 principal components; the sum of the explained variance percentage of the first 6 principal components is 99.5\%, suggesting that those principal components already contain most of the information from the original 35 persistent homology features in our dataset.

In Figure 4, we show the performance of GNN models (with global max pooling layers, see discussions below), including CGCNN, GAT, Transformer networks, on our constructed O-based perovskite dataset. By including the ASPH features, the MAE decreases for all GNN networks, suggesting that ASPH features can consistently improve the prediction accuracy on defect-related tasks. Since those features can be easily incorporated into the node features, we expect that the improvement will also be observed for other GNN models. Furthermore, among all networks, the Transformer network performs the best, with the MAE reduced by 55\% from 1.55 eV to 0.72 eV. This agrees with our previous findings in the configurationally disordered materials~\cite{Fang24configurational_GNN}. In Transformer networks, the features of the central node and adjacent nodes are treated as queries and keys, respectively. The overlap of the queries and keys reflects the similarity between the two nodes, determined by not only the elemental properties but also the chemical environment incorporated by ASPH features. Since the ASPH features entail the defect-related structure information such as the number of nearby defects and the distance to those defects, the attention mechanism allows the network to effectively extract the chemical information of defects and make accurate predictions on the defect properties. In addition to the above networks, we also trained the equivariant neural network (E3NN) on the dataset, and similar to other networks, introducing the ASPH features reduces the MAE from 1.94 eV to 1.27 eV (see Supplementary Materials II). On our dataset, E3NN network performs worse than the Transformer network, possibly due to the fact that both the node features and the prediction target are invariant quantities and also due to the diverse chemical bonding network in the host materials in our dataset. Nevertheless, we note that ASPH features can still improve the performance of the E3NN network.

\begin{figure}[htb]
\centering
\includegraphics[width=\linewidth]{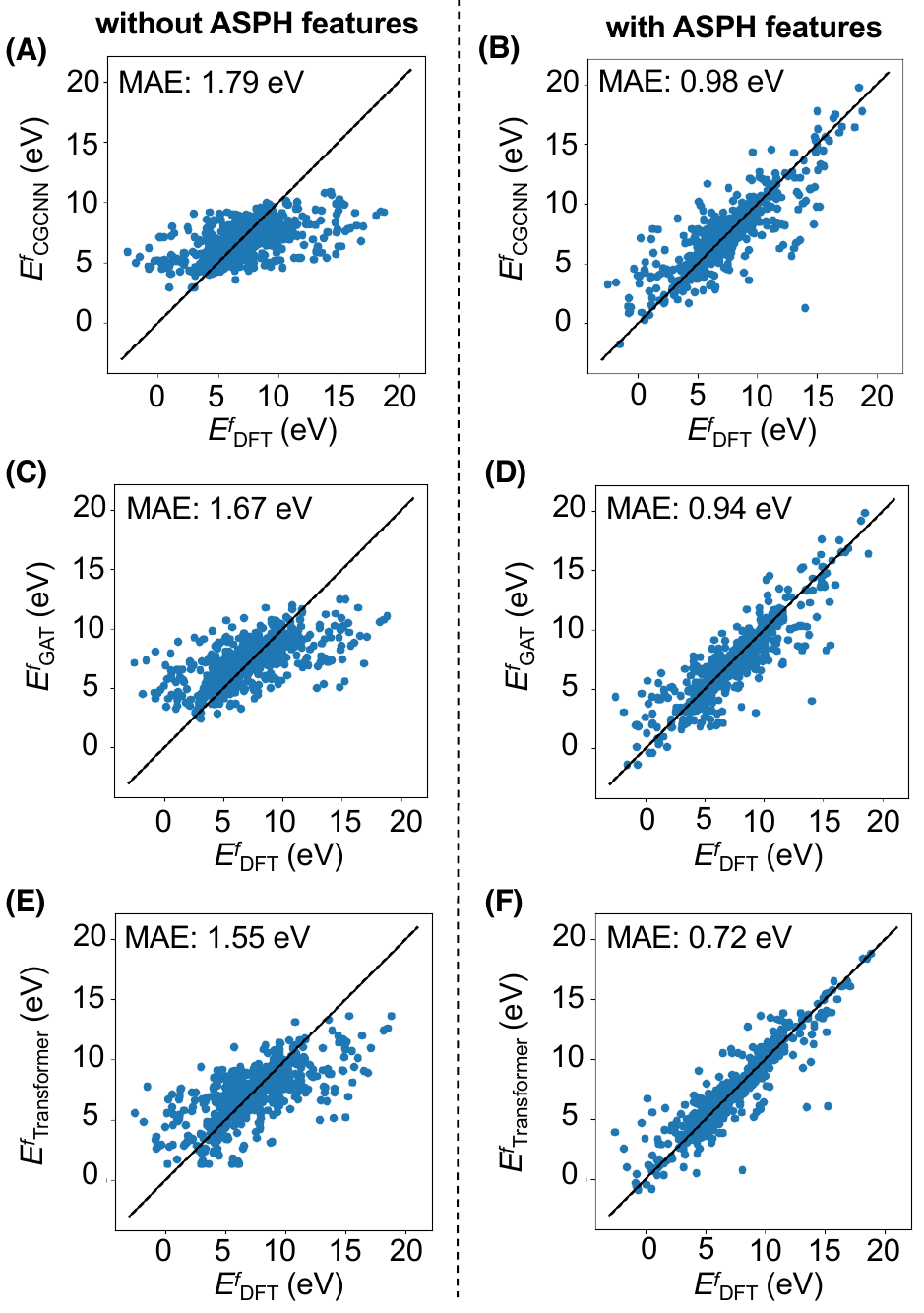}
\caption{The comparison of the performance of GNN models (using global max pooling layers) on the O-based perovskite dataset without and with ASPH features. (A, B) CGCNN networks. (C, D) GAT networks. (E, F) Transformer networks.}
\label{fig:MAEs}
\end{figure}


Furthermore, in addition to the node features, we note that graph pooling layers will also affect the performance of GNN models on defect-related tasks. Previous GNN models for material property predictions usually use global mean (sum) pooling layers to collect all node features, which calculates the mean (sum) of those features as the global representation of the graph. However, this method includes the contributions from all node features and thus averages out the distinct features related to defects, as discussed above. Therefore, instead of global mean (sum) pooling, we choose to use global max pooling layers to extract the maximum features over all nodes, which will direct the focus of our GNN models to the most distinct features of atoms near the defect sites. The MAEs using different pooling operations are summarized in Table~2 (more discussions can be found in Supplementary Materials II). Those results suggest that global max pooling layers are effective in extracting the defect information from the large supercell structures.

\begin{table*}[htb]
    \centering
    \begin{tabular}{|c|c|c|c|c|}
    \hline
    global pooling operations & max & sum & mean & min-max \\ \hline
    without ASPH features & 1.55 & 1.75 & 1.74 & 1.79 \\ \hline
    with ASPH features & 0.72 & 0.90 & 0.86 & 0.85 \\ \hline
    \end{tabular}
    \caption{The comparison of MAE of the Transformer models using different global pooling layers on the \ch{O}-based perovskite dataset without and with ASPH features.}
    \label{tab:GNN_pooling}
\end{table*}

More importantly, the global max pooling layers, along with the ASPH features, can help overcome the previously reported limitation where the performance of GNN models is not convergent with respect to the size of the defective supercells~\cite{Rahman24Defect_ALIGNN}. Here, we demonstrate that our GNN model which uses global max pooling layers can produce converged MAE with respect to the supercell size. Since the crystal systems of our host (pristine) perovskites are diverse and the defective supercells are constructed nearly cubic, making it computationally expensive to increase the periodicity in all directions, we choose to focus only on host perovskites in the cubic and the tetragonal crystal system, where the transformation matrix to the defective supercell is diagonal, so that increasing the periodicity along Cartesian directions does not change the overall shape of the supercell significantly. Based on the constructed defective supercells (which already contain hundreds of atoms), we consecutively increase the size of the supercell only in the direction along which the lattice constants are the smallest, as shown in Figure 5(A), while keeping the number of vacancies fixed, assuming that the defect formation energies from DFT methods remain constant. As the size of the supercell increases, the number of atoms within each supercell increases drastically, as shown in Figure 5(B). In panel (C), the MAE of the Transformer GNN model using the global mean pooling layers increases gradually as the size of the supercell increases due to the "trivial" atoms irrelevant to the defect property. On the other hand the MAEs of our current model using the global max pooling layers and the ASPH features are convergent with respect to the supercell size (though the MAE is larger than that trained on the whole dataset, possibly due to less amount of data), demonstrating that our GNN model can also adapt to the dilute defect limit.

\begin{figure}[htb]
\centering
\includegraphics[width=\linewidth]{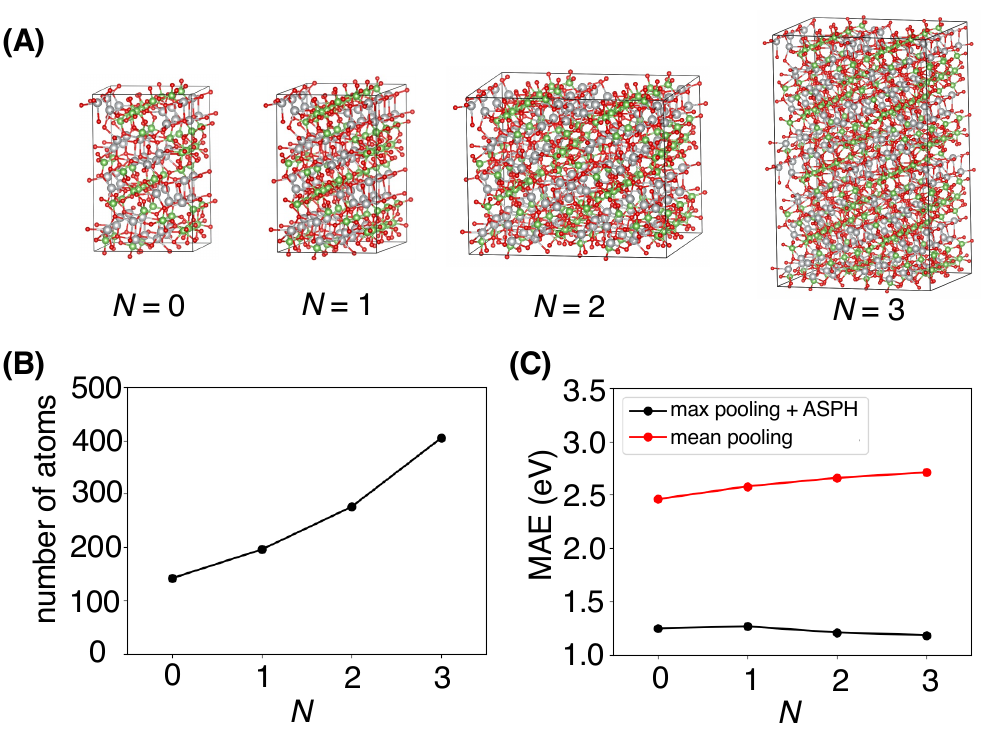}
\caption{(A) Illustration of the defective supercell with increasing sizes, where $N = 0$ is the original defective supercell, and $N > 0$ represents the supercell by repeating the $N-1$ cell along the directions along which the lattice constant is the smallest. (B) The average number of atoms of the defective supercells of different sizes. (C) The MAEs of the Transformer GNN model using global max pooling layers and ASPH features (black line), and using global mean pooling layers and without ASPH features (red line) on the dataset of defective supercells of different sizes.}
\label{fig:supercell}
\end{figure}

Finally, we narrow our dataset down to structures with only \ch{O}-vacancies. For the whole dataset, the accurate prediction of the vacancy formation energies requires information about the local chemical environment around the vacancy as well as the elemental type of the vacancy. As we focus on only \ch{O}-vacancies, the prediction task is simpler than that on the whole dataset since the elemental type information of the vacancy is no longer needed. However, most of the host materials in our dataset still possess multiple symmetry-inequivalent \ch{O}-sites with varying vacancy formation energies, making it challenging for previously proposed machine learning and GNN models to tackle. With our GNN model with ASPH features and global max pooling layers, the MAE is 0.45~eV, which is comparable to most previous machine learning and deep learning models predicting oxygen vacancy formation energies in oxides \cite{Wan21ML_O_vac, Park24ML_O_vac, Baldassarri23ML_O_vac}. 

\subsection{Multi-Defect Interaction Energy in \ch{BaTiO3}}
As discussed above in the toy models, persistent homology features contain information not only about vacancies, but also about substitution defects. Therefore, we anticipate that those features can also improve the prediction performance for structures with substitutions. Here we use the cubic perovskite \ch{BaTiO3} as an example and consider the defect formation energies of multiple substitutions. We construct the dataset containing 1000 defective \ch{BaTiO3} in a $3 \times 3 \times 3$ supercell, and randomly replace at most 8 \ch{Ba} (out of 27) atoms by \ch{Ca} or \ch{Sr} and at most 4 \ch{Ti} (out of 27) atoms replaced by \ch{Zr}; those substitutions are common in \ch{BaTiO3}~\cite{Zhao20BaTiO3}. By training our GNN model on this dataset, we find that after including the persistent homology features, the MAE decreases slightly from 0.07~eV~to 0.03~eV, and the comparison between defect formation energies of DFT and GNN methods are shown in Figure~6(a). We note that substitutional defects do not change the overall chemical bonding network of \ch{BaTiO3}, but only replace atoms by other chemical species randomly, similar to configurationally disordered materials~\cite{Fang24configurational_GNN}. In this case, even without the persistent homology features, our GNN model can already capture the elemental swaps and make accurate predictions on the formation energies, and adding the persistent homology features only slightly increases the accuracy. 

Furthermore, we consider \ch{BaTiO3} with multiple vacancies. The above discussions revealed that persistent homology features can also capture the number of vacancies around each atom. Therefore, persistent homology features can also benefit predictions with multiple vacancies. We construct another dataset containing 1000 defective \ch{BaTiO3} in a $3 \times 3 \times 3$ supercell, and randomly remove (exactly) 1 \ch{Ba}, 1 \ch{Ti}, and 3 \ch{O} atoms. Since we constrained the host material and also the number of vacancies, the energy fluctuations of different defect configurations arise from the vacancy-vacancy interactions. Therefore, the prediction target in fact reflects the multi-vacancy interaction energy; however, for consistency, we still call it the defect formation energy and calculate it from Equation~\eqref{eqn:defect_formation_energy}. By including the ASPH feature, the MAE decreases from 0.55~eV~to 0.43~eV, where the comparison between DFT and GNN formation energies is shown in Figure~6(b). The worse performance on tasks related to vacancy defects than to substitutional defects is because vacancies disrupt the chemical bonding network in materials and affect the exchange of information indirectly through the adjacency matrix of graphs, making it more challenging to capture the vacancy formation energies than substitution formation energies. Besides, by the construction of ASPH features (Figure~1), those features are naturally more sensitive to substitutions as exotic chemical species, while vacancies are obliquely encoded in the birth and death of 1-Betti numbers and 2-Betti numbers. We believe more direct encoding of the vacancy information in the persistent homology features, along with advanced GNN models which capture the interrupted chemical bonding network due to defects (such as higher-order GNNs), could further improve the prediction accuracy on defect properties.

\begin{figure}[htb]
\centering
\includegraphics[width=\linewidth]{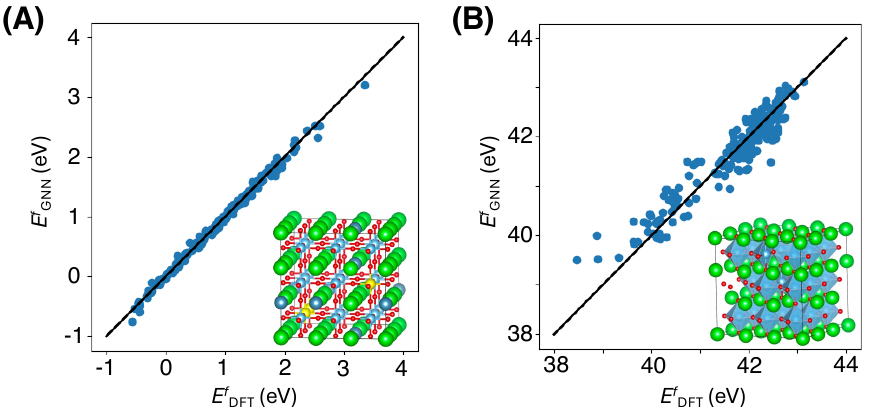}
\caption{(A) The comparison between the defect formation energies from the DFT method and the GNN model on the testing set of the \ch{BaTiO3} dataset with multiple substitutions. The inset is a typical structure in the dataset, with the following color code: light green: Ba; dark green: Sr; grey: Ca; yellow: Zr; cyan: Ti; red: O. (B) Same as (A), but for the dataset of \ch{BaTiO3} with multiple vacancies.}
\label{fig:BTO}
\end{figure}

\section{Conclusions}
In this work, we reveal the relationship between persistent homology features, obtained from the topological invariants of the homology group of point clouds, and the defect structures including multiple vacancies and substitutions. Through several toy models, we demonstrate that persistent homology features contain the local structural information near each atom, such as the type and the number of defects, the distance to those defects from each atom, and the size of the defects. These persistent homology features can be easily incorporated into the node features in any existing GNN models and improve the performance of GNN models in predicting defect-related properties. 

We numerically verify the role of persistent homology features using the dataset of O-based perovskite materials, which covers a wide spectrum of crystal systems and vacancies of various elemental types. We construct and train the GNN models on the dataset to predict the neutral vacancy formation energies. The results demonstrate that including the persistent homology features, along with global max pooling layers, can significantly increase the prediction accuracy of GNN models with the MAE reduced by around 55\%, outperforming previous machine learning and GNN models. Our model also produces MAE that does not depend on the supercell sizes, allowing for accurate predictions of defect properties in the low density limit.

Furthermore, we demonstrate that persistent homology features can help predict the defect-defect interaction energies using a dataset of the defective cubic perovskite BaTiO\textsubscript{3} with multiple substitutions and with multiple vacancies. These results justify the essential role of persistent homology features in the accurate predictions of defect properties, such as defect formation energies and charge transition levels, using machine learning and deep learning methods, and will benefit future studies on high-throughput screening and design of the combination of host materials and specific defects with desired properties and functionalities.

\section{Methods}
\subsection{Density Functional Theory Calculations}
First-principles calculations are carried out using the Vienna Ab initio Simulation Package~\cite{Kresse93VASP1, Kresse96VASP2} with projector augmented wave pseudopotentials~\cite{Kresse99PAW1, Blochl94PAW2}. We used the Perdew-Burke-Ernzerhof functional in the generalized gradient approximation for all calculations~\cite{Perdew96GGA}. The kinetic energy cutoff of the plane-wave basis sets is 550~eV, and the energy convergence threshold is $10^{-6}$~eV. The $\Gamma$ $\mathbf{k}$-point sampling scheme is used for all pristine and defective supercell calculations, and for elemental energy calculations we used the $\mathbf{k}$-point grid with the $\mathbf{k}$-point density of $0.03 (2\pi / \text{\AA})$. We performed the spin-polarized calculations with the DFT+U scheme for $3d$ elements, with the $U$ values taken from Materials Project~\cite{Jain13Materials_project}. The DFT-D3 method was also used to take account of the van der Waals interactions~\cite{Grimme10DFTD3, Grimme11DFTD3} in all calculations. 

In this work we only consider the neutral charge state for all defects, and therefore the defect formation energy for the defect $M_C$, following the Kr\"{o}ger–Vink notation, is defined as
\begin{equation}
    E^f [M_C] = E[M_C] - E[\text{bulk}] + \sum_i n_i \mu_i \label{eqn:defect_formation_energy}
\end{equation}
where $n_i$ is the difference in the number of elements between the defective and the pristine structures ($n_i > 0$ for removed atoms and $n_i < 0$ for added atoms), and $\mu_i$ is the chemical potential of element $i$, which we assumed to be the energy of the most stable single crystal of this element (except for \ch{H}, \ch{N}, \ch{O}, \ch{F}, \ch{Cl}, which we calculate their energies from corresponding gas molecules)~\cite{VdW04defect_review, Freysoldt14defect_review}.

\subsection{Training GNN Models}
To train our GNN model, our dataset is split into the training set, validation set, and testing set according to 60:20:20 ratio. The validation set is used to prevent overfitting on the training set, which usually occurs when the dataset contains few data. The MAE on the validation set is tracked during the training procedure, and we use the model with the minimal MAE on the validation set as our optimal model for testing.

The hyperparameters are optimized based on the Bayesian optimization method as implemented in Optuna~\cite{Optuna}, which calculates the expected improvement from a new set of hyperparameters using the tree-structured Parzen Estimator method \cite{Bergstra11TPE}, and decides whether to accept or decline this set of hyperparameters. The set of hyperparameters used in this work includes the number of convolution layers, the number of hidden channels in each convolution layer, the number of attention heads (used only in GNN models with the attention mechanism), learning rate, weight decay, and batch size. 

\section*{Data Availability}
The datasets and codes used in this work are available on Github at \url{https://github.com/qmatyanlab/Defect_GNN}.

\section*{Acknowledgements}
Z.F. thanks helpful advice from Alex Heilman. This work is supported by the National Science Foundation under Grant No. DMR-2314050. This research used resources of the National Energy Research Scientific Computing Center (NERSC), a U.S. Department of Energy Office of Science User Facility located at Lawrence Berkeley National Laboratory, operated under Contract No. DE-AC02-05CH11231 using NERSC award BESERCAP0029544.

\section*{Author contributions statement}

Z.F. and Q.Y. conceived the experiments. Z.F. wrote the code for graph neural networks and performed data analysis. Q.Y. supervised the study. The manuscript was written through contributions from all authors, and all authors have given approval to the final version of the manuscript.

\section*{Competing interests}
The authors declare no competing interests.

\section*{Additional information}
\subsection*{Supplementary Materials}
Supplementary Notes 1 to 2\\
Supplementary Table 1 to 3\\
Supplementary Figures 1 to 3\\

\bibliography{bibliography}

\providecommand{\latin}[1]{#1}
\makeatletter
\providecommand{\doi}
  {\begingroup\let\do\@makeother\dospecials
  \catcode`\{=1 \catcode`\}=2 \doi@aux}
\providecommand{\doi@aux}[1]{\endgroup\texttt{#1}}
\makeatother
\providecommand*\mcitethebibliography{\thebibliography}
\csname @ifundefined\endcsname{endmcitethebibliography}  {\let\endmcitethebibliography\endthebibliography}{}
\begin{mcitethebibliography}{60}
\providecommand*\natexlab[1]{#1}
\providecommand*\mciteSetBstSublistMode[1]{}
\providecommand*\mciteSetBstMaxWidthForm[2]{}
\providecommand*\mciteBstWouldAddEndPuncttrue
  {\def\EndOfBibitem{\unskip.}}
\providecommand*\mciteBstWouldAddEndPunctfalse
  {\let\EndOfBibitem\relax}
\providecommand*\mciteSetBstMidEndSepPunct[3]{}
\providecommand*\mciteSetBstSublistLabelBeginEnd[3]{}
\providecommand*\EndOfBibitem{}
\mciteSetBstSublistMode{f}
\mciteSetBstMaxWidthForm{subitem}{(\alph{mcitesubitemcount})}
\mciteSetBstSublistLabelBeginEnd
  {\mcitemaxwidthsubitemform\space}
  {\relax}
  {\relax}

\bibitem[Li \latin{et~al.}(2022)Li, Fu, Hu, and Peng]{Li22mechanical_review}
Li,~C.; Fu,~T.; Hu,~H.; Peng,~X. Mechanical properties and their sensitivity to point defects: $(\mathrm{HfNbTaTiZr})\mathrm{C}$ high-entropy carbide. \emph{Phys. Rev. B} \textbf{2022}, \emph{105}, 224102\relax
\mciteBstWouldAddEndPuncttrue
\mciteSetBstMidEndSepPunct{\mcitedefaultmidpunct}
{\mcitedefaultendpunct}{\mcitedefaultseppunct}\relax
\EndOfBibitem
\bibitem[Bostr\"{o}m and Kieslich(2021)Bostr\"{o}m, and Kieslich]{Bostrom21mechanical_review}
Bostr\"{o}m,~H. L.~B.; Kieslich,~G. Influence of Metal Defects on the Mechanical Properties of ABX3 Perovskite-Type Metal-formate Frameworks. \emph{J. Phys. Chem. C} \textbf{2021}, \emph{125}, 1467--1471\relax
\mciteBstWouldAddEndPuncttrue
\mciteSetBstMidEndSepPunct{\mcitedefaultmidpunct}
{\mcitedefaultendpunct}{\mcitedefaultseppunct}\relax
\EndOfBibitem
\bibitem[Dastider \latin{et~al.}(2023)Dastider, Rasul, Rahman, and Alam]{Dastider23mechanical_review}
Dastider,~A.~G.; Rasul,~A.; Rahman,~E.; Alam,~M.~K. Effect of vacancy defects on the electronic and mechanical properties of two-dimensional MoSi2N4. \emph{RSC Adv.} \textbf{2023}, \emph{13}, 5307--5316\relax
\mciteBstWouldAddEndPuncttrue
\mciteSetBstMidEndSepPunct{\mcitedefaultmidpunct}
{\mcitedefaultendpunct}{\mcitedefaultseppunct}\relax
\EndOfBibitem
\bibitem[Pan \latin{et~al.}(2013)Pan, Yang, Fu, Zhang, and Xu]{Pan13electronic_review}
Pan,~X.; Yang,~M.-Q.; Fu,~X.; Zhang,~N.; Xu,~Y.-J. Defective TiO<sub>2</sub> with oxygen vacancies: synthesis, properties and photocatalytic applications. \emph{NANOSCALE} \textbf{2013}, \emph{5}, 3601--3614\relax
\mciteBstWouldAddEndPuncttrue
\mciteSetBstMidEndSepPunct{\mcitedefaultmidpunct}
{\mcitedefaultendpunct}{\mcitedefaultseppunct}\relax
\EndOfBibitem
\bibitem[Yin \latin{et~al.}(2015)Yin, Yang, Kang, Yan, and Wei]{Yin15electronic_review}
Yin,~W.-J.; Yang,~J.-H.; Kang,~J.; Yan,~Y.; Wei,~S.-H. Halide perovskite materials for solar cells: a theoretical review. \emph{JOURNAL OF MATERIALS CHEMISTRY A} \textbf{2015}, \emph{3}, 8926--8942\relax
\mciteBstWouldAddEndPuncttrue
\mciteSetBstMidEndSepPunct{\mcitedefaultmidpunct}
{\mcitedefaultendpunct}{\mcitedefaultseppunct}\relax
\EndOfBibitem
\bibitem[Leem \latin{et~al.}(2024)Leem, Fang, Lee, Kim, Kakekhani, Liu, Cho, Kim, Wang, Ji, Patra, Kronik, Rappe, Yim, and Agarwal]{Leem24electronic_review}
Leem,~Y.-C.; Fang,~Z.; Lee,~Y.-K.; Kim,~N.-Y.; Kakekhani,~A.; Liu,~W.; Cho,~S.-P.; Kim,~C.; Wang,~Y.; Ji,~Z.; Patra,~A.; Kronik,~L.; Rappe,~A.~M.; Yim,~S.-Y.; Agarwal,~R. Optically Triggered Emergent Mesostructures in Monolayer WS2. \emph{Nano Letters} \textbf{2024}, \emph{24}, 5436--5443\relax
\mciteBstWouldAddEndPuncttrue
\mciteSetBstMidEndSepPunct{\mcitedefaultmidpunct}
{\mcitedefaultendpunct}{\mcitedefaultseppunct}\relax
\EndOfBibitem
\bibitem[Van~de Walle and Neugebauer(2004)Van~de Walle, and Neugebauer]{VdW04defect_review}
Van~de Walle,~C.~G.; Neugebauer,~J. {First-principles calculations for defects and impurities: Applications to III-nitrides}. \emph{Journal of Applied Physics} \textbf{2004}, \emph{95}, 3851--3879\relax
\mciteBstWouldAddEndPuncttrue
\mciteSetBstMidEndSepPunct{\mcitedefaultmidpunct}
{\mcitedefaultendpunct}{\mcitedefaultseppunct}\relax
\EndOfBibitem
\bibitem[Freysoldt \latin{et~al.}(2014)Freysoldt, Grabowski, Hickel, Neugebauer, Kresse, Janotti, and Van~de Walle]{Freysoldt14defect_review}
Freysoldt,~C.; Grabowski,~B.; Hickel,~T.; Neugebauer,~J.; Kresse,~G.; Janotti,~A.; Van~de Walle,~C.~G. First-principles calculations for point defects in solids. \emph{Rev. Mod. Phys.} \textbf{2014}, \emph{86}, 253--305\relax
\mciteBstWouldAddEndPuncttrue
\mciteSetBstMidEndSepPunct{\mcitedefaultmidpunct}
{\mcitedefaultendpunct}{\mcitedefaultseppunct}\relax
\EndOfBibitem
\bibitem[Oba \latin{et~al.}(2011)Oba, Choi, Togo, and Tanaka]{Oba11defect_migration}
Oba,~F.; Choi,~M.; Togo,~A.; Tanaka,~I. Point defects in ZnO: an approach from first principles. \emph{SCIENCE AND TECHNOLOGY OF ADVANCED MATERIALS} \textbf{2011}, \emph{12}, 3\relax
\mciteBstWouldAddEndPuncttrue
\mciteSetBstMidEndSepPunct{\mcitedefaultmidpunct}
{\mcitedefaultendpunct}{\mcitedefaultseppunct}\relax
\EndOfBibitem
\bibitem[Pochet and Caliste(2012)Pochet, and Caliste]{Pochet12defect_migration}
Pochet,~P.; Caliste,~D. Point defect diffusion in Si and SiGe revisited through atomistic simulations. \emph{MATERIALS SCIENCE IN SEMICONDUCTOR PROCESSING} \textbf{2012}, \emph{15}, 675--690\relax
\mciteBstWouldAddEndPuncttrue
\mciteSetBstMidEndSepPunct{\mcitedefaultmidpunct}
{\mcitedefaultendpunct}{\mcitedefaultseppunct}\relax
\EndOfBibitem
\bibitem[Lany and Zunger(2008)Lany, and Zunger]{Lany08defect_DFT}
Lany,~S.; Zunger,~A. Assessment of correction methods for the band-gap problem and for finite-size effects in supercell defect calculations: Case studies for ZnO and GaAs. \emph{Phys. Rev. B} \textbf{2008}, \emph{78}, 235104\relax
\mciteBstWouldAddEndPuncttrue
\mciteSetBstMidEndSepPunct{\mcitedefaultmidpunct}
{\mcitedefaultendpunct}{\mcitedefaultseppunct}\relax
\EndOfBibitem
\bibitem[Freysoldt and Neugebauer(2018)Freysoldt, and Neugebauer]{Freysoldt18defect_DFT}
Freysoldt,~C.; Neugebauer,~J. First-principles calculations for charged defects at surfaces, interfaces, and two-dimensional materials in the presence of electric fields. \emph{Phys. Rev. B} \textbf{2018}, \emph{97}, 205425\relax
\mciteBstWouldAddEndPuncttrue
\mciteSetBstMidEndSepPunct{\mcitedefaultmidpunct}
{\mcitedefaultendpunct}{\mcitedefaultseppunct}\relax
\EndOfBibitem
\bibitem[Frey \latin{et~al.}(2020)Frey, Akinwande, Jariwala, and Shenoy]{Frey20ML}
Frey,~N.~C.; Akinwande,~D.; Jariwala,~D.; Shenoy,~V.~B. Machine Learning-Enabled Design of Point Defects in 2D Materials for Quantum and Neuromorphic Information Processing. \emph{ACS Nano} \textbf{2020}, \emph{14}, 13406--13417\relax
\mciteBstWouldAddEndPuncttrue
\mciteSetBstMidEndSepPunct{\mcitedefaultmidpunct}
{\mcitedefaultendpunct}{\mcitedefaultseppunct}\relax
\EndOfBibitem
\bibitem[Deml \latin{et~al.}(2015)Deml, Holder, O’Hayre, Musgrave, and Stevanović]{Deml15ML}
Deml,~A.~M.; Holder,~A.~M.; O’Hayre,~R.~P.; Musgrave,~C.~B.; Stevanović,~V. Intrinsic Material Properties Dictating Oxygen Vacancy Formation Energetics in Metal Oxides. \emph{The Journal of Physical Chemistry Letters} \textbf{2015}, \emph{6}, 1948--1953\relax
\mciteBstWouldAddEndPuncttrue
\mciteSetBstMidEndSepPunct{\mcitedefaultmidpunct}
{\mcitedefaultendpunct}{\mcitedefaultseppunct}\relax
\EndOfBibitem
\bibitem[Wexler \latin{et~al.}(2021)Wexler, Gautam, Stechel, and Carter]{Wexler21ML}
Wexler,~R.~B.; Gautam,~G.~S.; Stechel,~E.~B.; Carter,~E.~A. Factors Governing Oxygen Vacancy Formation in Oxide Perovskites. \emph{Journal of the American Chemical Society} \textbf{2021}, \emph{143}, 13212--13227\relax
\mciteBstWouldAddEndPuncttrue
\mciteSetBstMidEndSepPunct{\mcitedefaultmidpunct}
{\mcitedefaultendpunct}{\mcitedefaultseppunct}\relax
\EndOfBibitem
\bibitem[Park \latin{et~al.}(2023)Park, Xu, Pan, Zhang, Lany, Liu, Luo, and Qi]{Park21ML}
Park,~J.; Xu,~B.; Pan,~J.; Zhang,~D.; Lany,~S.; Liu,~X.; Luo,~J.; Qi,~Y. Accurate prediction of oxygen vacancy concentration with disordered A-site cations in high-entropy perovskite oxides. \emph{npj Computational Materials} \textbf{2023}, \emph{9}, 29\relax
\mciteBstWouldAddEndPuncttrue
\mciteSetBstMidEndSepPunct{\mcitedefaultmidpunct}
{\mcitedefaultendpunct}{\mcitedefaultseppunct}\relax
\EndOfBibitem
\bibitem[Yan \latin{et~al.}(2024)Yan, Kar, Chowdhury, and Bansil]{Yan24ML}
Yan,~Q.; Kar,~S.; Chowdhury,~S.; Bansil,~A. The Case for a Defect Genome Initiative. \emph{Advanced Materials} \textbf{2024}, \emph{36}, 2303098\relax
\mciteBstWouldAddEndPuncttrue
\mciteSetBstMidEndSepPunct{\mcitedefaultmidpunct}
{\mcitedefaultendpunct}{\mcitedefaultseppunct}\relax
\EndOfBibitem
\bibitem[Chen \latin{et~al.}(2019)Chen, Ye, Zuo, Zheng, and Ong]{Chen19GNN}
Chen,~C.; Ye,~W.; Zuo,~Y.; Zheng,~C.; Ong,~S.~P. Graph Networks as a Universal Machine Learning Framework for Molecules and Crystals. \emph{Chem. Mater.} \textbf{2019}, \emph{31}, 3564--3572\relax
\mciteBstWouldAddEndPuncttrue
\mciteSetBstMidEndSepPunct{\mcitedefaultmidpunct}
{\mcitedefaultendpunct}{\mcitedefaultseppunct}\relax
\EndOfBibitem
\bibitem[Reiser \latin{et~al.}(2022)Reiser, Neubert, Eberhard, Torresi, Zhou, Shao, Metni, van Hoesel, Schopmans, Sommer, and Friederich]{Reiser22GNN}
Reiser,~P.; Neubert,~M.; Eberhard,~A.; Torresi,~L.; Zhou,~C.; Shao,~C.; Metni,~H.; van Hoesel,~C.; Schopmans,~H.; Sommer,~T.; Friederich,~P. Graph neural networks for materials science and chemistry. \emph{Commun. Mater.} \textbf{2022}, \emph{3}, 93\relax
\mciteBstWouldAddEndPuncttrue
\mciteSetBstMidEndSepPunct{\mcitedefaultmidpunct}
{\mcitedefaultendpunct}{\mcitedefaultseppunct}\relax
\EndOfBibitem
\bibitem[Fung \latin{et~al.}(2021)Fung, Zhang, Juarez, and Sumpter]{Fung21GNN}
Fung,~V.; Zhang,~J.; Juarez,~E.; Sumpter,~B.~G. Benchmarking graph neural networks for materials chemistry. \emph{npj Comput. Mater.} \textbf{2021}, \emph{7}, 84\relax
\mciteBstWouldAddEndPuncttrue
\mciteSetBstMidEndSepPunct{\mcitedefaultmidpunct}
{\mcitedefaultendpunct}{\mcitedefaultseppunct}\relax
\EndOfBibitem
\bibitem[Xie and Grossman(2018)Xie, and Grossman]{Xie18CGCNN}
Xie,~T.; Grossman,~J.~C. Crystal Graph Convolutional Neural Networks for an Accurate and Interpretable Prediction of Material Properties. \emph{Phys. Rev. Lett.} \textbf{2018}, \emph{120}, 145301\relax
\mciteBstWouldAddEndPuncttrue
\mciteSetBstMidEndSepPunct{\mcitedefaultmidpunct}
{\mcitedefaultendpunct}{\mcitedefaultseppunct}\relax
\EndOfBibitem
\bibitem[Kazeev \latin{et~al.}(2023)Kazeev, Al-Maeeni, Romanov, Faleev, Lukin, Tormasov, Castro~Neto, Novoselov, Huang, and Ustyuzhanin]{Kazeev23Defect_GNN}
Kazeev,~N.; Al-Maeeni,~A.~R.; Romanov,~I.; Faleev,~M.; Lukin,~R.; Tormasov,~A.; Castro~Neto,~A.~H.; Novoselov,~K.~S.; Huang,~P.; Ustyuzhanin,~A. Sparse representation for machine learning the properties of defects in 2D materials. \emph{npj Computational Materials} \textbf{2023}, \emph{9}, 113\relax
\mciteBstWouldAddEndPuncttrue
\mciteSetBstMidEndSepPunct{\mcitedefaultmidpunct}
{\mcitedefaultendpunct}{\mcitedefaultseppunct}\relax
\EndOfBibitem
\bibitem[Xiang \latin{et~al.}(2024)Xiang, Soh, and Dunham]{Xiang24Defect_GNN}
Xiang,~X.; Soh,~D.; Dunham,~S. Exploration of Deep Learning Models for Accelerated Defect Property Predictions and Device Design of Cubic Semiconductor Crystals. \emph{The Journal of Physical Chemistry C} \textbf{2024}, \emph{128}, 8821--8829\relax
\mciteBstWouldAddEndPuncttrue
\mciteSetBstMidEndSepPunct{\mcitedefaultmidpunct}
{\mcitedefaultendpunct}{\mcitedefaultseppunct}\relax
\EndOfBibitem
\bibitem[Rahman \latin{et~al.}(2024)Rahman, Gollapalli, Manganaris, Yadav, Pilania, DeCost, Choudhary, and Mannodi-Kanakkithodi]{Rahman24Defect_ALIGNN}
Rahman,~M.~H.; Gollapalli,~P.; Manganaris,~P.; Yadav,~S.~K.; Pilania,~G.; DeCost,~B.; Choudhary,~K.; Mannodi-Kanakkithodi,~A. {Accelerating defect predictions in semiconductors using graph neural networks}. \emph{APL Machine Learning} \textbf{2024}, \emph{2}, 016122\relax
\mciteBstWouldAddEndPuncttrue
\mciteSetBstMidEndSepPunct{\mcitedefaultmidpunct}
{\mcitedefaultendpunct}{\mcitedefaultseppunct}\relax
\EndOfBibitem
\bibitem[Li \latin{et~al.}(2018)Li, Han, and Wu]{Li18oversmoothing_GNN}
Li,~Q.; Han,~Z.; Wu,~X.-m. Deeper Insights Into Graph Convolutional Networks for Semi-Supervised Learning. \emph{Proceedings of the AAAI Conference on Artificial Intelligence} \textbf{2018}, \emph{32}, 1\relax
\mciteBstWouldAddEndPuncttrue
\mciteSetBstMidEndSepPunct{\mcitedefaultmidpunct}
{\mcitedefaultendpunct}{\mcitedefaultseppunct}\relax
\EndOfBibitem
\bibitem[Pun \latin{et~al.}(2022)Pun, Lee, and Xia]{Pun22persistent_homology}
Pun,~C.~S.; Lee,~S.~X.; Xia,~K. Persistent-homology-based machine learning: a survey and a comparative study. \emph{Artif. Intell. Rev.} \textbf{2022}, \emph{55}, 5169–5213\relax
\mciteBstWouldAddEndPuncttrue
\mciteSetBstMidEndSepPunct{\mcitedefaultmidpunct}
{\mcitedefaultendpunct}{\mcitedefaultseppunct}\relax
\EndOfBibitem
\bibitem[Wu \latin{et~al.}(2018)Wu, Zhao, Wang, and Wei]{Wu18persistent_homology}
Wu,~K.; Zhao,~Z.; Wang,~R.; Wei,~G.-W. TopP–S: Persistent homology-based multi-task deep neural networks for simultaneous predictions of partition coefficient and aqueous solubility. \emph{Journal of Computational Chemistry} \textbf{2018}, \emph{39}, 1444--1454\relax
\mciteBstWouldAddEndPuncttrue
\mciteSetBstMidEndSepPunct{\mcitedefaultmidpunct}
{\mcitedefaultendpunct}{\mcitedefaultseppunct}\relax
\EndOfBibitem
\bibitem[Cang and Wei(2017)Cang, and Wei]{Cang17persistent_homology}
Cang,~Z.; Wei,~G.-W. TopologyNet: Topology based deep convolutional and multi-task neural networks for biomolecular property predictions. \emph{PLOS Computational Biology} \textbf{2017}, \emph{13}, 1--27\relax
\mciteBstWouldAddEndPuncttrue
\mciteSetBstMidEndSepPunct{\mcitedefaultmidpunct}
{\mcitedefaultendpunct}{\mcitedefaultseppunct}\relax
\EndOfBibitem
\bibitem[Jiang \latin{et~al.}(2021)Jiang, Chen, Chen, Li, Wei, and Pan]{Jiang21persistent_homology}
Jiang,~Y.; Chen,~D.; Chen,~X.; Li,~T.; Wei,~G.-W.; Pan,~F. Topological representations of crystalline compounds for the machine-learning prediction of materials properties. \emph{npj Computational Materials} \textbf{2021}, \emph{7}, 28\relax
\mciteBstWouldAddEndPuncttrue
\mciteSetBstMidEndSepPunct{\mcitedefaultmidpunct}
{\mcitedefaultendpunct}{\mcitedefaultseppunct}\relax
\EndOfBibitem
\bibitem[Roedel \latin{et~al.}(2009)Roedel, Jo, Seifert, Anton, Granzow, and Damjanovic]{Roedel09BaTiO3}
Roedel,~J.; Jo,~W.; Seifert,~K. T.~P.; Anton,~E.-M.; Granzow,~T.; Damjanovic,~D. Perspective on the Development of Lead-free Piezoceramics. \emph{JOURNAL OF THE AMERICAN CERAMIC SOCIETY} \textbf{2009}, \emph{92}, 1153--1177\relax
\mciteBstWouldAddEndPuncttrue
\mciteSetBstMidEndSepPunct{\mcitedefaultmidpunct}
{\mcitedefaultendpunct}{\mcitedefaultseppunct}\relax
\EndOfBibitem
\bibitem[Dey and Wang(2022)Dey, and Wang]{dey22persistent_homology}
Dey,~T.~K.; Wang,~Y. \emph{Computational topology for data analysis}; Cambridge University Press, 2022\relax
\mciteBstWouldAddEndPuncttrue
\mciteSetBstMidEndSepPunct{\mcitedefaultmidpunct}
{\mcitedefaultendpunct}{\mcitedefaultseppunct}\relax
\EndOfBibitem
\bibitem[Aktas \latin{et~al.}(2019)Aktas, Akbas, and El~Fatmaoui]{Aktas19persistent_homology}
Aktas,~M.~E.; Akbas,~E.; El~Fatmaoui,~A. Persistence homology of networks: methods and applications. \emph{APPLIED NETWORK SCIENCE} \textbf{2019}, \emph{4}, 1\relax
\mciteBstWouldAddEndPuncttrue
\mciteSetBstMidEndSepPunct{\mcitedefaultmidpunct}
{\mcitedefaultendpunct}{\mcitedefaultseppunct}\relax
\EndOfBibitem
\bibitem[Zia \latin{et~al.}(2024)Zia, Khamis, Nichols, Tayab, Hayder, Rolland, Stone, and Petersson]{Zia24persistent_homology}
Zia,~A.; Khamis,~A.; Nichols,~J.; Tayab,~U.~B.; Hayder,~Z.; Rolland,~V.; Stone,~E.; Petersson,~L. Topological deep learning: a review of an emerging paradigm. \emph{ARTIFICIAL INTELLIGENCE REVIEW} \textbf{2024}, \emph{57}, 4\relax
\mciteBstWouldAddEndPuncttrue
\mciteSetBstMidEndSepPunct{\mcitedefaultmidpunct}
{\mcitedefaultendpunct}{\mcitedefaultseppunct}\relax
\EndOfBibitem
\bibitem[Fang and Yan(2024)Fang, and Yan]{Fang24configurational_GNN}
Fang,~Z.; Yan,~Q. Towards accurate prediction of configurational disorder properties in materials using graph neural networks. \emph{npj Computational Materials} \textbf{2024}, \emph{10}, 91\relax
\mciteBstWouldAddEndPuncttrue
\mciteSetBstMidEndSepPunct{\mcitedefaultmidpunct}
{\mcitedefaultendpunct}{\mcitedefaultseppunct}\relax
\EndOfBibitem
\bibitem[Jain \latin{et~al.}(2013)Jain, Ong, Hautier, Chen, Richards, Dacek, Cholia, Gunter, Skinner, Ceder, and Persson]{Jain13Materials_project}
Jain,~A.; Ong,~S.~P.; Hautier,~G.; Chen,~W.; Richards,~W.~D.; Dacek,~S.; Cholia,~S.; Gunter,~D.; Skinner,~D.; Ceder,~G.; Persson,~K.~A. {Commentary: The Materials Project: A materials genome approach to accelerating materials innovation}. \emph{APL Materials} \textbf{2013}, \emph{1}, 011002\relax
\mciteBstWouldAddEndPuncttrue
\mciteSetBstMidEndSepPunct{\mcitedefaultmidpunct}
{\mcitedefaultendpunct}{\mcitedefaultseppunct}\relax
\EndOfBibitem
\bibitem[Morris \latin{et~al.}(2019)Morris, Ritzert, Fey, Hamilton, Lenssen, Rattan, and Grohe]{Morris18higher_order_GNN}
Morris,~C.; Ritzert,~M.; Fey,~M.; Hamilton,~W.~L.; Lenssen,~J.~E.; Rattan,~G.; Grohe,~M. Weisfeiler and Leman Go Neural: Higher-Order Graph Neural Networks. \emph{Proceedings of the AAAI Conference on Artificial Intelligence} \textbf{2019}, \emph{33}, 4602--4609\relax
\mciteBstWouldAddEndPuncttrue
\mciteSetBstMidEndSepPunct{\mcitedefaultmidpunct}
{\mcitedefaultendpunct}{\mcitedefaultseppunct}\relax
\EndOfBibitem
\bibitem[Reiser \latin{et~al.}(2022)Reiser, Neubert, Eberhard, Torresi, Zhou, Shao, Metni, van Hoesel, Schopmans, Sommer, and Friederich]{Reiser22higher_order_GNN}
Reiser,~P.; Neubert,~M.; Eberhard,~A.; Torresi,~L.; Zhou,~C.; Shao,~C.; Metni,~H.; van Hoesel,~C.; Schopmans,~H.; Sommer,~T.; Friederich,~P. Graph neural networks for materials science and chemistry. \emph{COMMUNICATIONS MATERIALS} \textbf{2022}, \emph{3}, 1\relax
\mciteBstWouldAddEndPuncttrue
\mciteSetBstMidEndSepPunct{\mcitedefaultmidpunct}
{\mcitedefaultendpunct}{\mcitedefaultseppunct}\relax
\EndOfBibitem
\bibitem[Veličković \latin{et~al.}(2018)Veličković, Cucurull, Casanova, Romero, Liò, and Bengio]{GAT1}
Veličković,~P.; Cucurull,~G.; Casanova,~A.; Romero,~A.; Liò,~P.; Bengio,~Y. Graph Attention Networks. International Conference on Learning Representations. 2018\relax
\mciteBstWouldAddEndPuncttrue
\mciteSetBstMidEndSepPunct{\mcitedefaultmidpunct}
{\mcitedefaultendpunct}{\mcitedefaultseppunct}\relax
\EndOfBibitem
\bibitem[Brody \latin{et~al.}(2022)Brody, Alon, and Yahav]{GAT2}
Brody,~S.; Alon,~U.; Yahav,~E. How Attentive are Graph Attention Networks? International Conference on Learning Representations. 2022\relax
\mciteBstWouldAddEndPuncttrue
\mciteSetBstMidEndSepPunct{\mcitedefaultmidpunct}
{\mcitedefaultendpunct}{\mcitedefaultseppunct}\relax
\EndOfBibitem
\bibitem[Shi \latin{et~al.}(2021)Shi, Huang, Feng, Zhong, Wang, and Sun]{Shi21Transformer}
Shi,~Y.; Huang,~Z.; Feng,~S.; Zhong,~H.; Wang,~W.; Sun,~Y. Masked Label Prediction: Unified Message Passing Model for Semi-Supervised Classification. Proceedings of the Thirtieth International Joint Conference on Artificial Intelligence, {IJCAI-21}. 2021; pp 1548--1554\relax
\mciteBstWouldAddEndPuncttrue
\mciteSetBstMidEndSepPunct{\mcitedefaultmidpunct}
{\mcitedefaultendpunct}{\mcitedefaultseppunct}\relax
\EndOfBibitem
\bibitem[Zhou \latin{et~al.}(2020)Zhou, Cui, Hu, Zhang, Yang, Liu, Wang, Li, and Sun]{Zhou2020GNN_review}
Zhou,~J.; Cui,~G.; Hu,~S.; Zhang,~Z.; Yang,~C.; Liu,~Z.; Wang,~L.; Li,~C.; Sun,~M. Graph neural networks: A review of methods and applications. \emph{AI Open} \textbf{2020}, \emph{1}, 57--81\relax
\mciteBstWouldAddEndPuncttrue
\mciteSetBstMidEndSepPunct{\mcitedefaultmidpunct}
{\mcitedefaultendpunct}{\mcitedefaultseppunct}\relax
\EndOfBibitem
\bibitem[Batzner \latin{et~al.}(2022)Batzner, Musaelian, Sun, Geiger, Mailoa, Kornbluth, Molinari, Smidt, and Kozinsky]{Batzner22EquivariantGNN}
Batzner,~S.; Musaelian,~A.; Sun,~L.; Geiger,~M.; Mailoa,~J.~P.; Kornbluth,~M.; Molinari,~N.; Smidt,~T.~E.; Kozinsky,~B. E(3)-equivariant graph neural networks for data-efficient and accurate interatomic potentials. \emph{Nature communications} \textbf{2022}, \emph{13}, 2453--2453\relax
\mciteBstWouldAddEndPuncttrue
\mciteSetBstMidEndSepPunct{\mcitedefaultmidpunct}
{\mcitedefaultendpunct}{\mcitedefaultseppunct}\relax
\EndOfBibitem
\bibitem[Alkauskas \latin{et~al.}(2014)Alkauskas, Yan, and Van~de Walle]{Alkauskas14pymatgen_defect_analysis}
Alkauskas,~A.; Yan,~Q.; Van~de Walle,~C.~G. First-principles theory of nonradiative carrier capture via multiphonon emission. \emph{Phys. Rev. B} \textbf{2014}, \emph{90}, 075202\relax
\mciteBstWouldAddEndPuncttrue
\mciteSetBstMidEndSepPunct{\mcitedefaultmidpunct}
{\mcitedefaultendpunct}{\mcitedefaultseppunct}\relax
\EndOfBibitem
\bibitem[Freysoldt \latin{et~al.}(2009)Freysoldt, Neugebauer, and Van~de Walle]{Freysoldt09pymatgen_defect_analysis}
Freysoldt,~C.; Neugebauer,~J.; Van~de Walle,~C.~G. Fully Ab Initio Finite-Size Corrections for Charged-Defect Supercell Calculations. \emph{Phys. Rev. Lett.} \textbf{2009}, \emph{102}, 016402\relax
\mciteBstWouldAddEndPuncttrue
\mciteSetBstMidEndSepPunct{\mcitedefaultmidpunct}
{\mcitedefaultendpunct}{\mcitedefaultseppunct}\relax
\EndOfBibitem
\bibitem[Choudhary and DeCost(2021)Choudhary, and DeCost]{Choudhary2021ALIGNN}
Choudhary,~K.; DeCost,~B. Atomistic Line Graph Neural Network for improved materials property predictions. \emph{npj computational materials} \textbf{2021}, \emph{7}, 1--8\relax
\mciteBstWouldAddEndPuncttrue
\mciteSetBstMidEndSepPunct{\mcitedefaultmidpunct}
{\mcitedefaultendpunct}{\mcitedefaultseppunct}\relax
\EndOfBibitem
\bibitem[Greenacre \latin{et~al.}(2023)Greenacre, Groenen, Hastie, D’Enza, Markos, and Tuzhilina]{Greenacre2023PCA}
Greenacre,~M.; Groenen,~P. J.~F.; Hastie,~T.; D’Enza,~A.~I.; Markos,~A.; Tuzhilina,~E. Publisher Correction: Principal component analysis. \emph{Nature Reviews Methods Primers} \textbf{2023}, \emph{3}, 1\relax
\mciteBstWouldAddEndPuncttrue
\mciteSetBstMidEndSepPunct{\mcitedefaultmidpunct}
{\mcitedefaultendpunct}{\mcitedefaultseppunct}\relax
\EndOfBibitem
\bibitem[Wan \latin{et~al.}(2021)Wan, Wang, Liu, and Liang]{Wan21ML_O_vac}
Wan,~Z.; Wang,~Q.-D.; Liu,~D.; Liang,~J. Data-driven machine learning model for the prediction of oxygen vacancy formation energy of metal oxide materials. \emph{Physical chemistry chemical physics : PCCP} \textbf{2021}, \emph{23}, 15675--15684\relax
\mciteBstWouldAddEndPuncttrue
\mciteSetBstMidEndSepPunct{\mcitedefaultmidpunct}
{\mcitedefaultendpunct}{\mcitedefaultseppunct}\relax
\EndOfBibitem
\bibitem[Park \latin{et~al.}(2024)Park, Lee, Park, Park, Heo, and Lee]{Park24ML_O_vac}
Park,~S.; Lee,~N.; Park,~J.~O.; Park,~J.; Heo,~Y.~S.; Lee,~J. Exploring the Latent Chemical Space of Oxygen Vacancy Formation Energy by a Machine Learning Ensemble. \emph{ACS Materials Letters} \textbf{2024}, \emph{6}, 66--72\relax
\mciteBstWouldAddEndPuncttrue
\mciteSetBstMidEndSepPunct{\mcitedefaultmidpunct}
{\mcitedefaultendpunct}{\mcitedefaultseppunct}\relax
\EndOfBibitem
\bibitem[Baldassarri \latin{et~al.}(2023)Baldassarri, He, Gopakumar, Griesemer, Salgado-Casanova, Liu, Torrisi, and Wolverton]{Baldassarri23ML_O_vac}
Baldassarri,~B.; He,~J.; Gopakumar,~A.; Griesemer,~S.; Salgado-Casanova,~A. J.~A.; Liu,~T.-C.; Torrisi,~S.~B.; Wolverton,~C. Oxygen Vacancy Formation Energy in Metal Oxides: High-Throughput Computational Studies and Machine-Learning Predictions. \emph{Chemistry of Materials} \textbf{2023}, \emph{35}, 10619--10634\relax
\mciteBstWouldAddEndPuncttrue
\mciteSetBstMidEndSepPunct{\mcitedefaultmidpunct}
{\mcitedefaultendpunct}{\mcitedefaultseppunct}\relax
\EndOfBibitem
\bibitem[Zhao \latin{et~al.}(2020)Zhao, Huang, and Wu]{Zhao20BaTiO3}
Zhao,~C.; Huang,~Y.; Wu,~J. Multifunctional barium titanate ceramics via chemical modification tuning phase structure. \emph{InfoMat} \textbf{2020}, \emph{2}, 1163--1190\relax
\mciteBstWouldAddEndPuncttrue
\mciteSetBstMidEndSepPunct{\mcitedefaultmidpunct}
{\mcitedefaultendpunct}{\mcitedefaultseppunct}\relax
\EndOfBibitem
\bibitem[Kresse and Hafner(1993)Kresse, and Hafner]{Kresse93VASP1}
Kresse,~G.; Hafner,~J. Ab initio molecular dynamics for liquid metals. \emph{Phys. Rev. B} \textbf{1993}, \emph{47}, 558--561\relax
\mciteBstWouldAddEndPuncttrue
\mciteSetBstMidEndSepPunct{\mcitedefaultmidpunct}
{\mcitedefaultendpunct}{\mcitedefaultseppunct}\relax
\EndOfBibitem
\bibitem[Kresse and Furthmüller(1996)Kresse, and Furthmüller]{Kresse96VASP2}
Kresse,~G.; Furthmüller,~J. Efficiency of ab-initio total energy calculations for metals and semiconductors using a plane-wave basis set. \emph{Comput. Mater. Sci.} \textbf{1996}, \emph{6}, 15--50\relax
\mciteBstWouldAddEndPuncttrue
\mciteSetBstMidEndSepPunct{\mcitedefaultmidpunct}
{\mcitedefaultendpunct}{\mcitedefaultseppunct}\relax
\EndOfBibitem
\bibitem[Kresse and Joubert(1999)Kresse, and Joubert]{Kresse99PAW1}
Kresse,~G.; Joubert,~D. From ultrasoft pseudopotentials to the projector augmented-wave method. \emph{Phys. Rev. B} \textbf{1999}, \emph{59}, 1758--1775\relax
\mciteBstWouldAddEndPuncttrue
\mciteSetBstMidEndSepPunct{\mcitedefaultmidpunct}
{\mcitedefaultendpunct}{\mcitedefaultseppunct}\relax
\EndOfBibitem
\bibitem[Bl\"ochl(1994)]{Blochl94PAW2}
Bl\"ochl,~P.~E. Projector augmented-wave method. \emph{Phys. Rev. B} \textbf{1994}, \emph{50}, 17953--17979\relax
\mciteBstWouldAddEndPuncttrue
\mciteSetBstMidEndSepPunct{\mcitedefaultmidpunct}
{\mcitedefaultendpunct}{\mcitedefaultseppunct}\relax
\EndOfBibitem
\bibitem[Perdew \latin{et~al.}(1996)Perdew, Burke, and Ernzerhof]{Perdew96GGA}
Perdew,~J.~P.; Burke,~K.; Ernzerhof,~M. Generalized Gradient Approximation Made Simple. \emph{Phys. Rev. Lett.} \textbf{1996}, \emph{77}, 3865--3868\relax
\mciteBstWouldAddEndPuncttrue
\mciteSetBstMidEndSepPunct{\mcitedefaultmidpunct}
{\mcitedefaultendpunct}{\mcitedefaultseppunct}\relax
\EndOfBibitem
\bibitem[Grimme \latin{et~al.}(2010)Grimme, Antony, Ehrlich, and Krieg]{Grimme10DFTD3}
Grimme,~S.; Antony,~J.; Ehrlich,~S.; Krieg,~H. A consistent and accurate ab initio parametrization of density functional dispersion correction (DFT-D) for the 94 elements H-Pu. \emph{The Journal of chemical physics} \textbf{2010}, \emph{132}, 154104--154104--19\relax
\mciteBstWouldAddEndPuncttrue
\mciteSetBstMidEndSepPunct{\mcitedefaultmidpunct}
{\mcitedefaultendpunct}{\mcitedefaultseppunct}\relax
\EndOfBibitem
\bibitem[Grimme \latin{et~al.}(2011)Grimme, Ehrlich, and Goerigk]{Grimme11DFTD3}
Grimme,~S.; Ehrlich,~S.; Goerigk,~L. Effect of the damping function in dispersion corrected density functional theory. \emph{Journal of computational chemistry} \textbf{2011}, \emph{32}, 1456--1465\relax
\mciteBstWouldAddEndPuncttrue
\mciteSetBstMidEndSepPunct{\mcitedefaultmidpunct}
{\mcitedefaultendpunct}{\mcitedefaultseppunct}\relax
\EndOfBibitem
\bibitem[Akiba \latin{et~al.}(2019)Akiba, Sano, Yanase, Ohta, and Koyama]{Optuna}
Akiba,~T.; Sano,~S.; Yanase,~T.; Ohta,~T.; Koyama,~M. Optuna: A Next-Generation Hyperparameter Optimization Framework. Proceedings of the 25th ACM SIGKDD International Conference on Knowledge Discovery \& Data Mining. 2019; p 2623–2631\relax
\mciteBstWouldAddEndPuncttrue
\mciteSetBstMidEndSepPunct{\mcitedefaultmidpunct}
{\mcitedefaultendpunct}{\mcitedefaultseppunct}\relax
\EndOfBibitem
\bibitem[Bergstra \latin{et~al.}(2011)Bergstra, Bardenet, Bengio, and K\'{e}gl]{Bergstra11TPE}
Bergstra,~J.; Bardenet,~R.; Bengio,~Y.; K\'{e}gl,~B. Algorithms for Hyper-Parameter Optimization. Advances in Neural Information Processing Systems. 2011\relax
\mciteBstWouldAddEndPuncttrue
\mciteSetBstMidEndSepPunct{\mcitedefaultmidpunct}
{\mcitedefaultendpunct}{\mcitedefaultseppunct}\relax
\EndOfBibitem
\end{mcitethebibliography}

\begin{figure}[htb]
\centering
\includegraphics[width=3.25in]{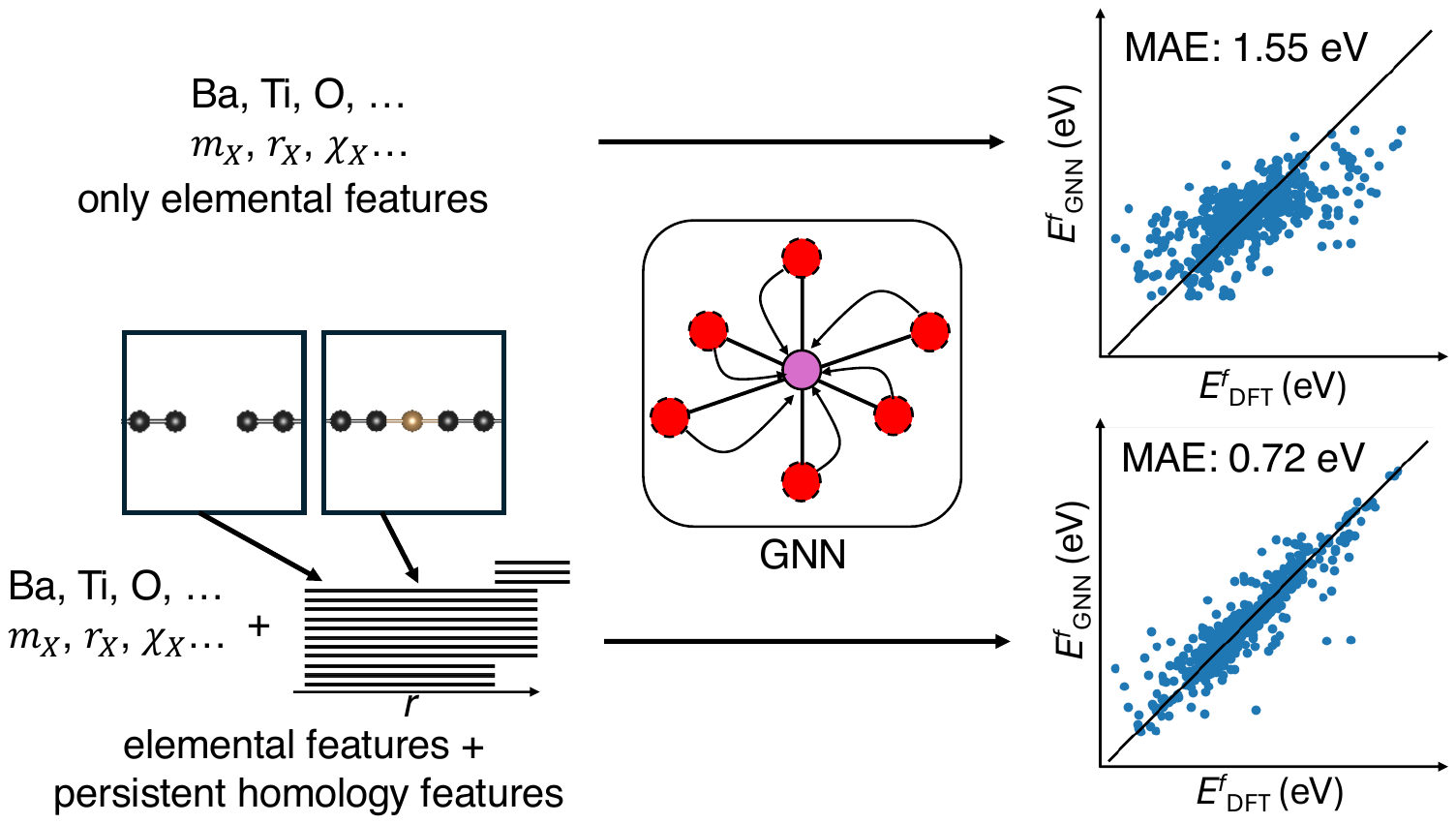}
\caption{For Table of Contents Only}
\end{figure}

\end{document}